\documentclass[showpacs,aps]{revtex4}

\usepackage{amsfonts}
\usepackage{amsmath}
\usepackage{amssymb}
\usepackage{graphicx}

\usepackage{graphicx}
\begin{document}

\baselineskip 17pt

\title{The angular spin current and its physical consequences}
\author{Qing-feng Sun$^{1,*}$ and X. C. Xie$^{2,3}$}
\affiliation{$^1$Beijing National Lab for Condensed Matter Physics
and Institute of Physics, Chinese Academy of Sciences, Beijing
100080, China \\
$^2$Department of Physics, Oklahoma State University, Stillwater,
Oklahoma 74078 \\
$^3$International Center for Quantum structures, Chinese Academy
of Sciences, Beijing 100080, China }
\date{\today}

\begin{abstract}
We find that in order to completely describe the spin transport,
apart from spin current (or linear spin current), one has to
introduce the angular spin current. The two
spin currents respectively describe the
translational and rotational motion of a spin. The definitions
of these spin current densities are given and their physical
properties are discussed. Both spin current densities
appear naturally in the spin continuity equation.
Moreover we predict that the angular spin current can
also induce an electric field $\vec{E}$, and in particular
$\vec{E}$ scales as $1/r^2$ at large distance $r$, whereas the
$\vec{E}$ field generated from the linear spin current goes as
$1/r^3$.
\end{abstract}


\maketitle

Recently, a new sub-discipline of condensed matter physics,
spintronics, is emerging rapidly and generating great
interests.\cite{ref1,ref2} The spin current, the most
important physical quantity in spintronics, has been extensively
studied. Many interesting and fundamental phenomena, such as the spin
Hall effect\cite{ref3,ref4,ref5} and the spin
precession\cite{ref6,ref7} in systems with spin-orbit
coupling, have been discovered and are under further study.

As for the charge current, the definition of the local charge
current density $\vec{j}^e({\bf r},t) = Re [\Psi^{\dagger}({\bf
r},t) \vec{v} \Psi({\bf r},t)]$ and its continuity equation
$
 \frac{d}{dt} \rho ({\bf r},t) + \nabla \bullet \vec{j}^e ({\bf r},t) =0
$ is well-known in physics. Here $\Psi({\bf r},t)$ is the
electronic wave function, $\vec{v} = \dot{\bf r}$ is the velocity
operator, and $\rho({\bf r},t)=\Psi^{\dagger}\Psi$ is the charge
density. This continuity equation is the consequence of charge
invariance, i.e. when an electron moves from one place to another,
its charge remains the same. However, in the spin transport, there
are still a lot of debates over what is the correct definition for
spin current. The problem stems from the lack of understanding of
the spin continuity equation. Recently, some studies have begun
investigation in this direction, e.g. a semi-classical description
of the spin continuity equation has been proposed\cite{niu,zhang}.

In this paper, we study the quantum version of the continuity
equation and the definition of local spin current density. We find
that in order to completely describe the spin transport, one has
to use two quantities, the linear spin current and the angular
spin current, which respectively describe the translational and
rotational motion of a spin. The conventional linear spin current
has been extensively studied. However, the angular spin current
has not been investigated before. The definition of two spin
current densities are given and they appear naturally in the
quantum spin continuity equation. Moreover, we predict that the
angular spin current can generate an electric field similar as
with the linear spin current, and thus contains physical
consequences.

Before studying the spin current in a quantum system, we first
consider the classical case. Consider a classical particle having
a vector $\vec{m}$ (e.g. the classical magnetic moment, etc.) with
its magnitude $|\vec{m}|$ fixed under the particle motion.
To completely describe this vector flow (see Fig.1a), we
need three quantities: the local vector density $\vec{M}({\bf
r},t) =\rho({\bf r},t) \vec{m}({\bf r},t)$, the velocity
$\vec{v}({\bf r},t)$, and the angular velocity $\vec{\omega}({\bf
r},t)$. Here $\rho({\bf r},t)$ is the particle density, and
$\vec{v}$ and $\vec{\omega}$ describe the translational and rotational
motion, respectively. Since $|\vec{m}|$ is a constant, the
continuity equation is (see Fig.1a):
$
 \frac{d}{dt}\iiint_V \vec{M} dV =
  -\oint_S d\vec{S} \bullet \vec{v}\vec{M}
  + \iiint_V \vec{\omega} \times \vec{M} dV
$, or it can be rewritten in the differential form:
\begin{equation}
 \frac{d}{dt} \vec{M}({\bf r},t) =
  - \nabla \bullet\vec{v}({\bf r},t)\vec{M}({\bf r},t)
  + \vec{\omega}({\bf r},t)\times \vec{M}({\bf r},t),
\end{equation}
where $\vec{v}\vec{M}$ is a tensor, and its element
$(\vec{v}\vec{M})_{ij}=v_i M_j$.

Now we study the electronic spin $\vec{s}$ in the quantum case.
Consider an arbitrary wave function $\Psi({\bf r},t)$. The local
spin density $\vec{s}$ at the position ${\bf r}$ and time $t$ is:
$\vec{s}({\bf r},t)= \Psi^{\dagger}({\bf r},t) \hat{\vec{s}}
\Psi({\bf r},t)$, where $\hat{\vec{s}} =
\frac{\hbar}{2}\hat{\vec{\sigma}}$ with $\hat{\vec{\sigma}}$ being the
Pauli matrices. The time-derivative of $\vec{s}({\bf r},t)$ is:
\begin{equation}
 \frac{d}{dt}\vec{s}({\bf r},t) = \frac{\hbar}{2}
  \left\{
    \left[\frac{d}{dt} \Psi^{\dagger} \right]
    \hat{\vec{\sigma}} \Psi
   + \Psi^{\dagger} \hat{\vec{\sigma}}
    \frac{d}{dt} \Psi \right\}.
\end{equation}
From the Schrodinger equation, we have $\frac{d}{dt}\Psi({\bf
r},t)=\frac{1}{i\hbar}H \Psi({\bf r},t)$ and
$\frac{d}{dt}\Psi^{\dagger}({\bf r},t)=\frac{1}{-i\hbar}[H
\Psi({\bf r},t)]^{\dagger}$. Notice here the transposition in the
symbol $\dagger$ only acts on the spin indexes. By using the above
two equations, the Eq.(2) changes into:
\begin{equation}
  (d/dt) \vec{s}({\bf r},t) =
  [\Psi^{\dagger}\hat{\vec{\sigma}} H\Psi
   - (H\Psi)^{\dagger} \hat{\vec{\sigma}} \Psi ]/2i.
\end{equation}
In the derivation below, we use the following Hamiltonian:
\begin{equation}
  H= \frac{\vec{p}^2}{2m} +V({\bf r}) + \hat{\vec{\sigma}}\bullet
  \vec{B} +\frac{\alpha}{\hbar} \hat{z}\bullet
  (\hat{\vec{\sigma}}\times \vec{p}).
\end{equation}
Note that our results are independent of this specific choice of
the Hamiltonian.\cite{note1} In Eq.(4) the 1st and 2nd terms are
the kinetic energy and potential energy. The 3rd term is the
Zeeman energy due to a magnetic field, and the last term is the
Rashba spin-orbit coupling\cite{ref8,ref9}, which has been
extensively studied recently\cite{ref4,ref6,ref7}. Next we
substitute the Hamiltonian Eq.(4) into Eq.(3), and calculate the
right side of Eq.(3) one term by one term:

(1). For the 1st term of Eq.(4), $\frac{\vec{p}^2}{2m}$:
\begin{eqnarray}
 &&
 \mbox{}\hspace{-8mm}
  \Psi^{\dagger} \hat{\vec{\sigma}} (\vec{p}^2/2m)\Psi -
    [(\vec{p}^2/2m) \Psi]^{\dagger} \hat{\vec{\sigma}} \Psi
    \nonumber \\
 && =  - (\hbar^2/2m) \{\Psi^{\dagger}
   \hat{\vec{\sigma}} \nabla\bullet\nabla  \Psi -
    [\nabla\bullet\nabla \Psi]^{\dagger} \hat{\vec{\sigma}} \Psi
    \}
    \nonumber \\
 && =  -(\hbar^2/2m) \nabla\bullet \{ \Psi^{\dagger}
   \nabla \hat{\vec{\sigma}} \Psi -
    [\nabla \Psi]^{\dagger} \hat{\vec{\sigma}} \Psi \}
    \nonumber \\
 && = -i\hbar \nabla \bullet Re \{ \Psi^{\dagger}
  (\vec{p}/m) \hat{\vec{\sigma}} \Psi \}
\end{eqnarray}
Here the dotted multiply between the vector and the tensor is
defined as usually, e.g. $(\vec{A} \bullet {\bf B})_j \equiv
\sum_i A_i B_{ij}$ with the vector $\vec{A}$ and the tensor ${\bf
B}$.

(2). For the 2nd term of Eq.(4), $V({\bf r})$: its corresponding
value is zero.

(3). For the 3rd term, $\hat{\vec{\sigma}}\bullet \vec{B}$:
\begin{eqnarray}
 && \mbox{}\hspace{-8mm}
 \Psi^{\dagger}\hat{\vec{\sigma}}(\hat{\vec{\sigma}}\bullet\vec{B})\Psi
  -[\hat{\vec{\sigma}}\bullet\vec{B}\Psi]^{\dagger}\hat{\vec{\sigma}}\Psi
   \nonumber\\
  && =  \Psi^{\dagger}(\hat{\vec{\sigma}}\hat{\vec{\sigma}}\bullet
   \vec{B}
   -\vec{B}\bullet\hat{\vec{\sigma}}\hat{\vec{\sigma}})\Psi
   =2i\Psi^{\dagger}\vec{B}\times\hat{\vec{\sigma}}\Psi
\end{eqnarray}

(4). For the last term,
$\frac{\alpha}{\hbar}\hat{z}\bullet(\hat{\vec{\sigma}}\times\vec{p})
=\frac{\alpha}{\hbar}\hat{\vec{\sigma}}\bullet(\vec{p}\times\hat{z})$:
\begin{eqnarray}
 && \mbox{}\hspace{-2mm}
 \Psi^{\dagger}\hat{\vec{\sigma}}\hat{\vec{\sigma}}\bullet(\vec{p}\times\hat{z})
  \Psi
  -[\hat{\vec{\sigma}}\bullet(\vec{p}\times\hat{z})\Psi]^{\dagger}\hat{\vec{\sigma}}\Psi
   \nonumber\\
&& =
 \Psi^{\dagger}(\vec{p}\times\hat{z}) \bullet \hat{\vec{\sigma}}\hat{\vec{\sigma}}
  \Psi
   +\Psi^{\dagger}2i(\vec{p}\times\hat{z})\times\hat{\vec{\sigma}}\Psi
  -[\hat{\vec{\sigma}}\bullet(\vec{p}\times\hat{z})\Psi]^{\dagger}\hat{\vec{\sigma}}\Psi
   \nonumber\\
 && =  -i\hbar\{
   \Psi^{\dagger}(\nabla\bullet(\hat{z}\times \hat{\vec{\sigma}})
    \hat{\vec{\sigma}}  \Psi
    +[\nabla\Psi]^{\dagger}\bullet (\hat{z}\times \hat{\vec{\sigma}})
    \hat{\vec{\sigma}}  \Psi \}
  +2i\Psi^{\dagger}(\vec{p}\times\hat{z})\times\hat{\vec{\sigma}}\Psi
   \nonumber\\
 && =
 -i\hbar\nabla\bullet\{\Psi^{\dagger}(\hat{z}\times\hat{\vec{\sigma}})
 \hat{\vec{\sigma}}\Psi\}
   +2i\Psi^{\dagger}(\vec{p}\times\hat{z})\times\hat{\vec{\sigma}}\Psi
   \nonumber
\end{eqnarray}
In the similar method, one also has:
\begin{eqnarray}
 && \mbox{}\hspace{-10mm}
 \Psi^{\dagger}\hat{\vec{\sigma}}\hat{\vec{\sigma}}\bullet(\vec{p}\times\hat{z})
  \Psi
  -[\hat{\vec{\sigma}}\bullet(\vec{p}\times\hat{z})\Psi]^{\dagger}\hat{\vec{\sigma}}\Psi
   \nonumber\\
 & & \mbox{}\hspace{-8mm}  =
 -i\hbar\nabla\bullet\{\Psi^{\dagger}(\hat{z}\times\hat{\vec{\sigma}})
 \hat{\vec{\sigma}}\Psi\}^*
   +2i [(\vec{p}\times\hat{z})\times\hat{\vec{\sigma}} \Psi]^{\dagger}
   \Psi  \nonumber
\end{eqnarray}
Therefore, we obtain the result of the last term:
\begin{eqnarray}
 && \mbox{}\hspace{-9mm}
  (\alpha/\hbar)\{\Psi^{\dagger}\hat{\vec{\sigma}}\hat{\vec{\sigma}}\bullet(\vec{p}\times\hat{z})
  \Psi
  -[\hat{\vec{\sigma}}\bullet(\vec{p}\times\hat{z})\Psi]^{\dagger}\hat{\vec{\sigma}}\Psi \}
   \nonumber\\
&& \mbox{}\hspace{-8mm} =
 -i\alpha\nabla\bullet Re \Psi^{\dagger}(\hat{z}\times\hat{\vec{\sigma}})
 \hat{\vec{\sigma}}\Psi
   + \frac{2i\alpha}{\hbar} Re \Psi^{\dagger}(\vec{p}\times\hat{z})\times\hat{\vec{\sigma}}\Psi
\end{eqnarray}

To summarize the above four terms, the Eq.(3) changes into:
\begin{eqnarray}
  \frac{d}{dt}\vec{s} & = & -\frac{\hbar}{2}\nabla\bullet
   Re\{\Psi^{\dagger} [\frac{\vec{p}}{m}
   +\frac{\alpha}{\hbar}(\hat{z}\times\hat{\vec{\sigma}})]
   \hat{\vec{\sigma}}\Psi \} \nonumber\\
 & +& Re\{ \Psi^{\dagger} [\vec{B}
  +\frac{\alpha}{\hbar}\vec{p}\times\hat{z}]\times\hat{\vec{\sigma}}
  \Psi \}.
\end{eqnarray}
Introducing a tensor ${\bf j}_s({\bf r},t)$ and a vector
$\vec{j}_{\omega}({\bf r},t)$:
\begin{eqnarray}
  {\bf j}_s({\bf r},t) &=& Re\{\Psi^{\dagger}
   [\frac{\vec{p}}{m}+\frac{\alpha}{\hbar}(\hat{z}\times\hat{\vec{\sigma}})]
    \hat{\vec{s}}\Psi \}\\
  \vec{j}_{\omega}({\bf r},t) &=& Re \{\Psi^{\dagger}
    \frac{2}{\hbar} [\vec{B} +
    \frac{\alpha}{\hbar}(\vec{p}\times\hat{z})]\times\hat{\vec{s}}\Psi
    \},
\end{eqnarray}
then Eq.(3) reduces to:
\begin{eqnarray}
  \frac{d}{dt} \vec{s}({\bf r},t) =
   -\nabla \bullet {\bf j}_s({\bf  r}, t)
    + \vec{j}_{\omega}({\bf r}, t),
\end{eqnarray}
or it can also be rewritten in the integral form:
\begin{equation}
  \frac{d}{dt} \iiint_V \vec{s} dV =
   - \oint_S d\vec{S} \bullet {\bf j}_s
    + \iiint_V \vec{j}_{\omega} dV.
\end{equation}
Due to the fact that $\hat{\vec{v}}=\frac{d}{dt}{\bf r}
=\frac{\vec{p}}{m} +\frac{\alpha}{\hbar}(\hat{z}\times
\hat{\vec{\sigma}})$ and $\frac{d}{dt}\hat{\vec{\sigma}}
=\frac{1}{i\hbar}[\hat{\vec{\sigma}},H] =\frac{2}{\hbar}[\vec{B} +
\frac{\alpha}{\hbar} \vec{p}\times\hat{z}]\times
\hat{\vec{\sigma}}$, Eqs.(9) and (10) become:
\begin{eqnarray}
  {\bf j}_s({\bf r},t) &=& Re\{\Psi^{\dagger}({\bf r},t)
    \hat{\vec{v}} \hat{\vec{s}}\Psi({\bf r},t) \} \\
  \vec{j}_{\omega}({\bf r},t) &=& Re\{\Psi^{\dagger}
    (d\hat{\vec{s}}/dt) \Psi \} = Re\{ \Psi^{\dagger}
    \hat{\vec{\omega}} \times \hat{\vec{s}} \Psi \},
\end{eqnarray}
where $\hat{\vec{\omega}}\equiv \frac{2}{\hbar} [\vec{B} +
\frac{\alpha}{\hbar}(\vec{p}\times\hat{z})]$.\cite{note1}

Eq.(11) is the spin continuity equation, which is very similar
with the classic vector continuity equation (1). In fact, this
spin continuity equation (11) is the consequence of invariance of
the spin magnitude $|\vec{s}|$, i.e. when an electron makes a
motion, either translation or rotation, its spin magnitude
$|\vec{s}|=\frac{\hbar}{2}$ remains a constant. The two quantities
${\bf j}_s({\bf r},t)$ and $\vec{j}_{\omega}({\bf r},t)$ in
Eq.(11), which are defined in Eqs.(13,14) respectively, describe
the translational and rotational motion of a spin at the location
${\bf r}$ and the time $t$. They will be named the linear and the
angular spin current densities accordingly\cite{add1note1}. In
fact, the linear spin current ${\bf j}_s({\bf r},t)$ is identical
with the conventional spin current investigated in recent
studies\cite{note2}. However, the angular spin current
$\vec{j}_{\omega}({\bf r},t)$ is new and has not been investigated
before.\cite{add1note1}

Next, we discuss certain properties of ${\bf j}_s({\bf r}, t)$ and
$\vec{j}_{\omega}({\bf r},t)$. Notice that $\vec{j}_{\omega}({\bf
r},t)$ which describe the rotational motion of the spin plays a
parallel role in comparison with the conventional linear spin
current ${\bf j}_s({\bf r}, t)$ for the spin transport. (1) The
linear spin current is a tensor. Its element, e.g. $j_{s,xy}$,
represents an electron moving along the $x$ direction with its
spin in the $y$ direction (see Fig.1b). The angular spin current
$\vec{j}_{\omega}$ is a vector. In Fig.1c, its element
$j_{\omega,x}$ describes the rotational motion of the spin in the
$y$ direction and the angular velocity ${\vec {\omega}}$ in the
$-z$ direction. (2) Why is it necessary to introduce two
quantities instead of one to describe a spin motion? The reason is
the same as for completely describing the motion of a classic
vector or a rigid body. There two quantities are used: the linear
velocity for the translational motion and the angular velocity for
the rotation. It is similar for the present quantum case, both
${\bf j}_s({\bf r},t)$ and $\vec{j}_{\omega}({\bf r},t)$ are
required to describe the motion of a quantum spin. (3) From the
linear spin current density ${\bf j}_s({\bf r}, t)$, one can
calculate (or say how much) the linear spin current $\vec{I}_s$
flowing through a surface $S$ (see Fig.1e): $ \vec{I}_s^S =\iint_S
d\vec{S}\bullet {\bf j}_s$. However, the behavior for the angular
spin current is different. From the density $\vec{j}_{\omega}({\bf
r}, t)$, it is meaningless to determine how much the angular spin
current flowing through a surface $S$, because the angular spin
current describes the rotational motion not the movement. On the
other hand, one can calculate the total angular spin current
$\vec{I}_{\omega}^V$ in a volume $V$ from $\vec{j}_{\omega}$:
$\vec{I}_{\omega}^V = \iiint_V \vec{j}_{\omega}({\bf r},t) dV$.
(4) If the system is in a steady state, ${\bf j}_s$ and
$\vec{j}_{\omega}$ are independent of the time $t$, and
$\frac{d}{dt} \vec{s}({\bf r},t)=0$. Then the spin continuity
equation (11) reduces to: $ \nabla \bullet {\bf j}_s =
\vec{j}_{\omega} $ or $  \oint_S d\vec{S} \bullet {\bf j}_s =
\iiint_V \vec{j}_{\omega} dV $. This means that the total linear
spin current flowing out of a closed surface is equal to the total
angular spin current enclosed. If to further consider a quasi one
dimensional (1D) system (see Fig.1e), then one has $\vec{I}_s^{S'}
- \vec{I}_s^{S} =\vec{I}_{\omega}^{V}$. (5) It is easy to prove
that the spin currents in the present definitions of Eqs.(13,14)
are invariant under a space coordinate transformation. (6) The
linear spin current density ${\bf j}_s=Re\{\Psi^{\dagger} \vec{v}
\hat{\vec{s}} \Psi\}$ gives both the spin direction and the
direction of spin movement, so it completely describes the
translational motion. However, the angular spin current density,
$\vec{j}_{\omega}=Re\{\Psi^{\dagger}\frac{d\hat{\vec{s}}}{dt}\Psi
\} =Re\{\Psi^{\dagger} \vec{\omega}\times\hat{\vec{s}} \Psi \}$
involves the vector product of $\vec{\omega}\times\hat{\vec{s}}$,
not the tensor $\vec{\omega}\hat{\vec{s}}$. Is it correct or
sufficient to describe the rotational motion? For example, the
rotational motion of Fig.1c with the spin $\vec{s}$ in the $y$
direction and the angular velocity $\vec{\omega}$ in the $-z$
direction is different from the one in Fig.1d in which $\vec{s}$
is in the $z$ direction and $\vec{\omega}$ is in the $y$
direction, but their angular spin currents are completely the
same. Shall we distinguish them? It turns out that the physical
results produced by the above two rotational motions (Fig.1c and
1d) are indeed the same. For instance, the induced electric field
by them is identical since a spin $\vec{s}$ has only the direction
but no size (see detail discussion below). Thus, the vector
$\vec{j}_{\omega}$ is sufficient to describe the rotational
motion, and no tensor is necessary.

Recently, theoretic studies have suggested that the (linear) spin
current can induce an electric field $\vec{E}$\cite{ref10,ref11}.
Can the angular spin current also induce an electric field? If so
this gives a way of detecting the angular spin current. Following,
we study this question by using the method of equivalent magnetic
charge\cite{griffiths}. Let us consider a steady-state angular
spin current element $\vec{j}_{\omega} dV$ at the origin.
Associated with the spin $\vec{s}$, there is a magnetic moment
(MM) $\vec{m}=g\mu_B \vec{\sigma} = \frac{2g\mu_B}{\hbar} \vec{s}$
where $\mu_B$ is the Bohr magneton. Thus, corresponding to
$\vec{j}_{\omega}$, there is also a angular MM current
$\vec{j}_{m\omega} dV = \frac{2g\mu_B}{\hbar} \vec{j}_{\omega}
dV$. From above discussions, we already know that
$\vec{j}_{m\omega}$ (or $\vec{j}_{\omega}$) comes from the
rotational motion of a MM $\vec{m}$ (or $\vec{s}$) (see Fig.1c and
1d), and $\vec{j}_{m\omega} = \vec{\omega}\times \vec{m}$ (or
$\vec{j}_{\omega} = \vec{\omega}\times \vec{s}$). Under the method
of equivalent magnetic charge, the MM $\vec{m}$ is equivalent to
two magnetic charges: one with magnetic charge $+q$ located at
$\delta \hat{n}_m$ and the other with $-q$ at $-\delta \hat{n}_m$
(see Fig.1f). $\hat{n}_m$ is the unit vector of $\vec{m}$ and
$\delta$ is a tiny length. The angular MM current
$\vec{j}_{m\omega}$ is equivalent to two magnetic charge currents:
one is $\vec{j}_{+q}=\hat{n}_j q\delta |\vec{\omega}| \sin \theta
$ at the location $\delta \hat{n}_m$, the other is
$\vec{j}_{-q}=\hat{n}_j q\delta |\vec{\omega}| \sin \theta$ at
$-\delta \hat{n}_m$ (see Fig.1f), with $\hat{n}_j$ being the unit
vector of $\vec{j}_{m\omega}$ and $\theta$ the angle between
$\vec{\omega}$ and $\vec{m}$. In our previous work,\cite{ref11} we
have given the formulae of the electric field induced by a
magnetic charge current. The electric field induced by
$\vec{j}_{m\omega}dV$ can be calculated by adding the
contributions from the two magnetic charge currents. Let $\delta
\rightarrow 0$, and note that  $2q\delta \rightarrow |\vec{m}|$
and $|\vec{\omega}| |\vec{m}| \sin\theta =|\vec{j}_{m\omega}|$, we
obtain the electric field $\vec{E}_{\omega}$ generated by an
element of the angular spin current $\vec{j}_{\omega}dV$:
\begin{equation}
 \vec{E}_{\omega} = \frac{-\mu_0}{4\pi}\int \frac{  \vec{j}_{m\omega} dV \times
 {\bf r} }{r^3}
     = \frac{-\mu_0 g\mu_B}{h} \int \vec{j}_{\omega} dV \times
  \frac{ {\bf r} }{r^3}
\end{equation}
We also rewrite the electric field $\vec{E}_s$ generated by
an element of the linear spin current using
the tensor ${\bf j}_s $:\cite{ref11}
\begin{equation}
 \vec{E}_s
     = \frac{-\mu_0 g\mu_B}{h} \nabla \times \int {\bf j}_s dV
     \bullet
  \frac{ {\bf r} }{r^3},
\end{equation}
Below we emphasize three points: (i) In the large $r$ case, the
electric field $\vec{E}_{\omega}$ decays as $1/r^2$. Note that the
field from a linear spin current $\vec{E}_s$ goes as $1/r^3$. In
fact, in terms of generating an electric field, the angular spin
current is as effective as a magnetic charge current. (ii) In the
steady-state case, the total electric field $\vec{E}_T=
\vec{E}_{\omega}+\vec{E}_s$ contains the property: $\oint_C
\vec{E}_T \bullet d\vec{l} =0$, where $C$ is an arbitrary close
contour not passing through the region of spin current. However,
for each $\vec{E}_{\omega}$ or $\vec{E}_s$, $\oint_C
\vec{E}_{\omega}\bullet d\vec{l}$ or $\oint_C \vec{E}_s\bullet
d\vec{l}$ can be non-zero. (iii) As mentioned above, a angular
spin current $\vec{j}_{\omega}$ may consist of different
$\vec{\omega}$ and $\vec{s}$ (see Fig.1c and 1d). However, the
resulting electric field only depends on
$\vec{j}_{\omega}=\vec{\omega} \times \vec{s}$. This is because a
spin vector contains only a direction and a magnitude, but not a
spatial size (i.e. the distance $\delta$ approaches to zero). In
the limit $\delta \rightarrow 0$, both magnetic charge currents
$\vec{j}_{\pm q}$ reduce to $\vec{\omega} \times \vec{m}/2$ at the
origin. Therefore, the overall effect of the rotational motion is
only related to $\vec{\omega} \times \vec{m}$, not separately on
$\vec{\omega}$ and $\vec{m}$. Hence it is enough to describe the
spin rotational motion by using a vector $\vec{\omega} \times
\vec{s}$, instead of a tensor $\vec{\omega} \vec{s}$.

Now we give an example of applying the above formulas,
Eqs.(13,14), to calculate the spin currents and the electric
fields generated by them. Let us consider a quasi 1D quantum wire
having the Rashba spin orbit coupling, and its Hamiltonian is:
\begin{equation}
 H =  \frac{\vec{p}^2 }{2 m} + V(y,z) + \frac{\alpha}{\hbar}
 \sigma_z p_x  ,
\end{equation}
Here the other Rashba term $-\frac{\alpha}{\hbar} \sigma_x
p_z$ is neglected because z-direction is quantized\cite{ref6}.
Let $\Psi$ be a stationary wave function
\begin{eqnarray}
 \Psi({\bf r}) = \frac{\sqrt{2}}{2}
   \left( \begin{array}{l}
         e^{i(k-k_R)x} \\
         e^{i(k+k_R)x}
    \end{array} \right) \varphi(y,z),
\end{eqnarray}
where $k_R \equiv \alpha m/\hbar^2$ and $\varphi(y,z)$ is the
bound state wave function in the confined $y$ and $z$ directions.
$\Psi({\bf r})$ represents the spin precession in the $x$-$y$
plane while moving along the $x$ axis and this wave function was
discussed in recent studies.\cite{ref6,ref7} Using Eqs.(13,14),
the spin current densities of the wave function $\Psi({\bf r})$
are easily obtained. There are only two non-zero elements of ${\bf
j}_s({\bf r})$:
\begin{eqnarray}
  j_{sxx}({\bf r}) & = & (\hbar^2 k/2m)
  |\Psi(y,z)|^2  \cos 2k_R x ,\\
  j_{sxy}({\bf r}) &= & (\hbar^2 k /2m)
  |\Psi(y,z)|^2  \sin 2k_R x .
\end{eqnarray}
The non-zero elements of $\vec{j}_{\omega}({\bf r})$ are:
\begin{eqnarray}
  j_{\omega x}({\bf r}) & = & - ( \hbar^2 k k_R /m)
  |\Psi(y,z)|^2  \sin 2k_R x, \\
  j_{\omega y}({\bf r}) &= & (\hbar^2 k k_R /m)
  |\Psi(y,z)|^2 \cos 2k_R x.
\end{eqnarray}
Those spin current densities confirm with the intuitive picture of
an electron motion, precession in the $x$-$y$ plane and movement
in the $x$ direction (see Fig.2a). Moreover, they indeed satisfy
$-\nabla\bullet {\bf j}_s({\bf r}) +\vec{j}_{\omega}({\bf r})=0$.

Let us calculate the induced electric fields at the location ${\bf
r}=(x,y,z)$ by the above spin currents. Substituting the spin
currents of Eqs.(19-22) into Eqs.(15,16) and assuming the
transverse sizes of the 1D wire are much smaller than
$\sqrt{y^2+z^2}$, the induced fields $\vec{E}_{\omega}$ and
$\vec{E}_s$ can be obtained as:
\begin{eqnarray}
 &&\vec{E}_{\omega}
     = -a K_1(\beta) \sin{\tilde{x}}
      \left(\frac{\tilde{z} \cos \tilde{x}}{\beta\sin\tilde{x}} ,
 \frac{\tilde{z} }{\beta} ,
 -\frac{\tilde{y} }{\beta}  -
 \frac{K_0(\beta)}{K_1(\beta)}
 \right), \nonumber\\
&&\vec{E}_s =a \frac{K_1}{\beta} \sin\tilde{x}
 \left(0, \frac{2\tilde{z}\tilde{y}}{\beta^2}
  +\tilde{z}F(\beta),
  \frac{\tilde{z}^2 -\tilde{y}^2}{\beta^2}
  -\tilde{y}F(\beta) \right),\nonumber
\end{eqnarray}
where $\tilde{i}=2k_R i$ ($i=x,y,z$),
$\beta=\sqrt{\tilde{y}^2+\tilde{z}^2}$, $F(\beta)
=1+\frac{\tilde{y}K_0(\beta)}{\beta K_1(\beta)}$, the constant $a=
2\mu_0 g\mu_B \hbar k k_R^2 \rho_s/\pi m$, $\rho_s$ is the linear
density of moving electrons under the bias of an external voltage,
and $K_0$ and $K_1$ are the Bessel
functions. The total electric field is: $\vec{E}_T
=\vec{E}_{\omega}+\vec{E}_s = -\frac{a}{2k_R}\nabla \{
\frac{\tilde{z} \sin \tilde{x}}{\beta} K_1(\beta) \}$. The total
electric field $\vec{E}_T$ represents the one generated by a 1D
wire of electric dipole moment $\vec{p}_e = (0,0, c\sin\tilde{x})$
at the $x$ axis (see Fig.2b), where $c$ is a constant. It is
obvious that $\nabla \times \vec{E}_T =0$, i.e. $\oint_C \vec{E}_T
\bullet d\vec{l} =0$. However, in general $\oint_C
\vec{E}_{\omega}\bullet d\vec{l}$ and $\oint_C \vec{E}_s\bullet
d\vec{l}$ are separately non-zero. Finally, we estimate the
magnitude of $\vec{E}_T$. We use parameters consistent with
realistic experimental samples. Take the Rashba parameter $\alpha
= 3 \times 10^{-11}$eVm (corresponding to $k_R$=1/100nm for
$m=0.036m_e$), $\rho_s=10^6$/m (i.e. one moving electron per 1000
nm in length), and $k=k_F=10^8$/m. The electric potential
difference between the two points A and B (see Fig.2b) is about
$0.01\mu V$, where the positions of A and B are
$\frac{1}{2k_R}(\frac{\pi}{2},0,0.01)$ and
$\frac{1}{2k_R}(\frac{3\pi}{2},0,0.01)$. This value of the
potential is measurable with today's technology. Furthermore, with
the above parameters the electric field $\vec{E}_T$ at A or B is
about $10$V/m which is rather large.

In summary, we have introduced the angular spin current and its
role in the spin continuity equation. We point out that the
angular spin current can also induce an electric field and its
$\vec{E}$ field scales as  $1/r^2$ at large $r$.

{\bf Acknowledgments:} We gratefully acknowledge financial support
from the Chinese Academy of Sciences and NSFC under Grant No.
90303016 and No. 10474125. XCX is supported by US-DOE under Grant
No. DE-FG02-04ER46124 and NSF-MRSEC under DMR-0080054.

\newpage
\begin{figure}

\caption{(Color online) (a) Schematic diagram for a classic vector
flow. (b) The linear spin current element $j_{s,xy}$. (c) and (d)
The angular spin current element $j_{\omega,x}$. (e) The spin
current in a quasi 1D quantum wire. (f) The currents of two
magnetic charges that are equivalent to a angular MM current.
}\label{fig:1}

\caption{(Color online) (a) Schematic diagram for the spin precession
in the $x$-$y$ plane while moving along the $x$ axis. (b) A
1D wire of electric dipole moment $\vec{p}_e$. This configuration will
generate an electric field equivalent to the field
from the spin currents in (a).}\label{fig:2}

\end{figure}


\begin{references}
\bibitem{add1note1}
$\vec{j}_{\omega}$ is called the spin torque in the work\cite{niu}
in which the semi-classical approach to the spin continuity
equation is discussed. Here we consider that $\vec{j}_{\omega}$
describes the rotational motion of a spin, thus named the angular
spin current.
\end{references}
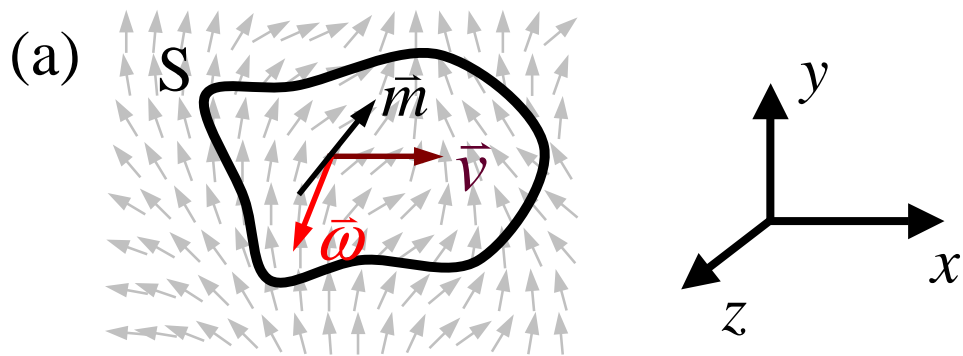
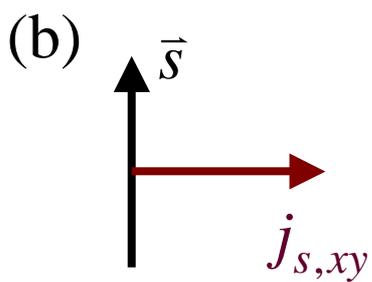
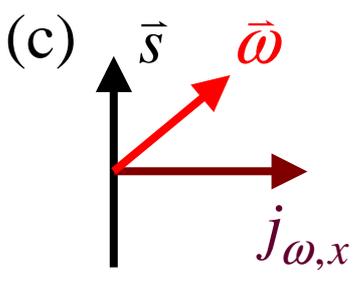
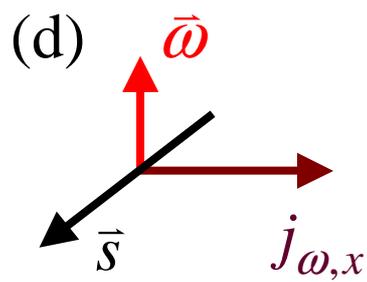
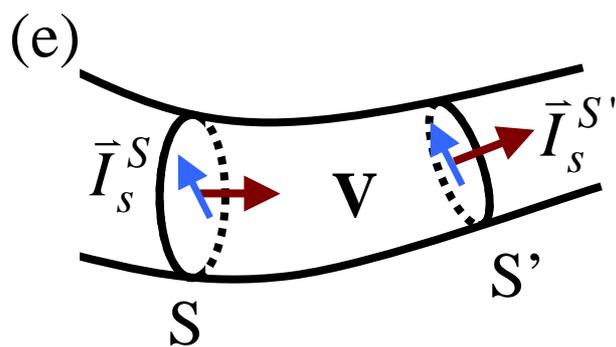
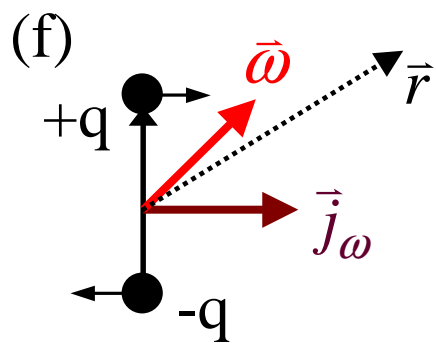
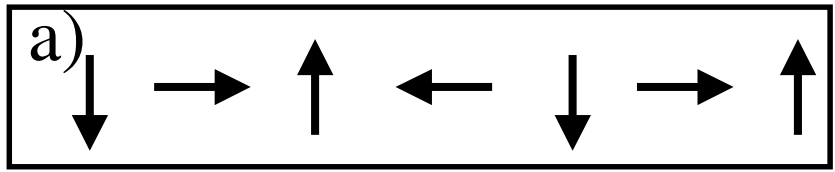
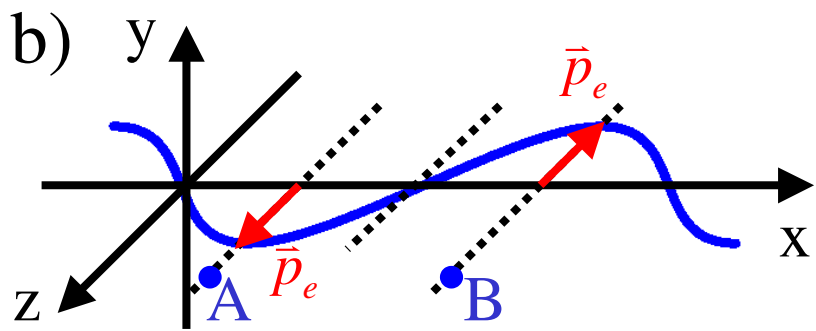
\end{document}